# Extremely terahertz electric-field enhancement in a high-Q photonic crystal slab cavity with nanoholes

QIJING LU,[1] XIAOGANG CHEN,[1] CHANG-LING ZOU,[2,3] AND SHUSEN XIE[1,4]

[1]*Key Laboratory of Optoelectronic Science and Technology for Medicine of Ministry of Education, Provincial Key Laboratory for Photonics Technology, Institute of Laser and Optoelectronics Technology, Fujian Normal University, Fuzhou 350007, China*
[2]*Key Laboratory of Quantum Information, Chinese Academy of Sciences, School of Physics, University of Science and Technology of China, Hefei 230026, P. R. China*
[3]*clzou321@ustc.edu.cn*
[4]*ssxie@fjnu.edu.cn*

**Abstract:** A one-dimensional photonic-crystal (PC) cavity with nanoholes is proposed for extremely enhancing the THz electric fields by utilizing the electromagnetic (EM) boundary conditions, where both slot effect (for the perpendicular component of the electric displacement field) and anti-slot effect (for the parallel component of the electric field) contribute to the considerable field enhancement. The EM energy density can be enhanced in the high refractive index material by a factor of $(\varepsilon_h/\varepsilon_l)^2$, where $\varepsilon_h$ and $\varepsilon_l$ are the permittivities of the high and low refractive index materials, respectively. Correspondingly, the mode volume can be enhanced by a factor of 288 as compared with the regular THz PC cavity and is three orders of magnitude smaller than the diffraction limitation. While the proposed THz cavity design also supports the modes with high $Q > 10^4$, which lead to strong Purcell enhancement of spontaneous emission by a factor exceeds $10^6$. Our THz cavity design is feasible and appealing for experimental demonstrations, since the semiconductor layer where the EM is maximized can naturally be filled with a quantum engineered active materials, which is potential for developing the room-temperature coherent THz radiation sources.

## 1. Introduction

Terahertz (THz) radiation is typically referred to electromagnetic waves of frequencies ranging from 0.1 to 10 THz, where 1 THz = $10^{12}$ Hz. Recently, tremendous progresses of THz technology have been achieved, since the THz waves are able to find applications in biological spectroscopy, label-free biosensing, imaging and security screening [1-4]. Efficiently generating coherent THz radiation [5, 6] is of particular importance for these applications. However, the radiative emission rate of a quantum emitter in the THz frequency band is limited due to small electric dipole moment [7]. The Purcell effect opens up a way to manipulate and enhance the rate of spontaneous emission of the emitter by modifying locally the electromagnetic (EM) density of states (DOS) for photons in microcavity [8]. Purcell factor [9], the enhancement factor of the spontaneous emission rate in a microcavity, scales with the ratio of quality ($Q$) factor and mode volume ($V$) [10]. Therefore, confining EM field into deep subwavelength volumes far beyond the diffraction limit of light can greatly modify DOS and enhance the Purcell factor. THz plasmonic microcavities based on metal-dielectric structure have been proposed [8, 11, 12], which were proved to support highly localized resonant modes concentrated in subwavelength mode volumes. Unfortunately, metal absorption loss makes the use of plasmon for building compact practical devices inadequate either in THz frequency band or in the optical domain. For example, THz plasmonic microcavities usually exhibit $Q$ factors below $10^2$ [13], and the Purcell factors are typically limited to below 50 [7, 8].

Very recently, all-dielectric THz whispering-gallery-mode (WGM) resonators [14-16] with high $Q$ factor of $10^3$-$10^4$ are exploited. WGM is known for high-$Q$, as the field is evenly distributed in the equator of the cavity via total internal reflection, however the mode volume is relatively large comparable to the wavelength to avoid the radiation loss. Besides plasmon and WGM-based devices, all-dielectric nanostructures have been utilized to enhance the local field for optical waves, such as the slot [17-19] and bowtie structures [20-23]. The enhancement is realized by the EM boundary conditions on the normal component of the electric displacement field [17-19] or parallel component of the electric field [22, 23]. However, at THz frequency range, the potential of all-dielectric structures to greatly enhance DOS has remained unexplored.

In this paper, we propose an all-dielectric semiconductor photonic-crystal (PC) slab cavity with nanoscale air holes and systematically explore the EM enhancement mechanisms in the THz frequency band. By using EM boundary conditions twice, the electric displacement field and electric field are both greatly enhanced by a factor of $\varepsilon_h/\varepsilon_l$, where $\varepsilon_h$ and $\varepsilon_l$ are the permittivities of the high and low refractive index materials; EM energy density is greatly enhanced by a factor of $(\varepsilon_h/\varepsilon_l)^2$. The THz radiation is squeezed into a mode volume that be three orders of magnitude smaller than the diffraction limited mode volume. With this excellent mode confinement, the radiation loss of the mode is suppressed and the $Q$ factor of resonant mode exceeds $10^4$, which is limited by the material absorption loss. The DOS in the proposed microcavity is greatly modified and the spontaneous emission rate of an emitter is enhanced by a Purcell factor of $\sim 10^6$. Compared to the slot or bowtie structures with EM field concentrated in the low refractive index materials, the proposed structure maximizes the EM field in the high refractive index materials, i.e., inside the cavity material. The proposed PC slab is monolithic and compatible with the current semiconductor fabrication technique. Thus, the proposed THz microcavity is particularly appealing for the realization of the coherent THz source in future.

## 2. Design and results

In the following, the properties of the proposed design are systematically studied by numerical simulation, with the eigenmodes of the THz cavity be solved by three-dimensional (3-D) finite-element method using the COMSOL Multiphysics 4.3. Figure 1(a) shows the schematic of the proposed two types of THz PC cavities with nanoholes placed in the center of the regular PC slab cavities. Firstly, we describe the structure of the regular THz PC cavities. It consists of one suspended semiconductor (Si) waveguide beam patterned with a 1-D line of air holes. The refractive index and extinction coefficient of Si are $n_{Si} = 3.42$ and $k_{Si} = 2.4\times10^{-5}$ [24, 25]. The width and thickness of Si waveguide are 50 μm and 22 μm, respectively. The hole-to-hole distance $a = 33$ μm is a constant. The hole-to-hole distance between the two central holes is also equal to the rest of the distance ($a$), i.e., there is no additional cavity length between the two Gaussian shape mirrors ($L = 0$). Quadratic radius-modulated air holes are used according to the deterministic cavity design process [26, 27]. Quadratic radius-modulated air holes are able to obtain Gaussian attenuated mirrors from the center to the end and such Gaussian shape mirrors could minimize the radiation loss [27, 28]. The $j^{th}$ ($j$ increases from 0 to $j_{max}$) hole's radius is defined as $r(j) = r_{center}+j^2(r_{end}-r_{center})/j_{max}^2$, where $r_{center} = 12.127$ μm and $r_{end} = 8.5749$ μm are the center and the end holes of the Gaussian mirrors in tapered region. This parabolic tapered air holes forms Bragg mirrors and Gaussian type confinement can be achieved [28]. Here, the $j_{max}$ is set as 20 and the hole radius in periodic mirror is set as $r_{end}$ in both side. The corresponding electric field distribution is plotted in Fig. 1(b), as expected, the THz wave is well confined in the cavity. The resonant frequency is 2.176 THz and the resonant mode is transverse-electric (TE) polarized (the electric field is in the $y$ direction). The Gaussian envelope of the 1-D distributions of $|E|$ is attributed from the parabolic tapered air holes.

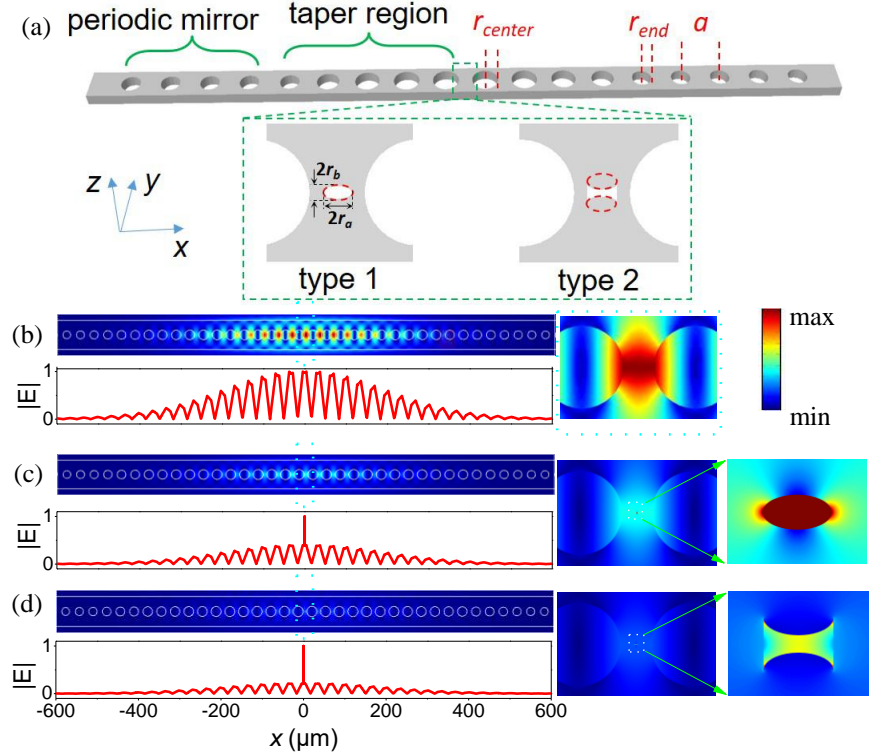

Fig. 1 (a) Schematic of the proposed THz PC cavities with nanoholes. In type-1 structure: an ellipse air hole with convex surface is inserted into the center of the cavity; In type-2 structure: two silicon holes shift in the opposite direction of the y-axis and an air hole with concave surface is left. 2-D and 1-D distributions of |E| in regular (b), type-1 (c) and type-2 (d) THz PC cavities. Right panels show the enlarged view of the field distributions in the center of the regular, type-1 and type-2 cavities and air holes in type-1 and type-2 cavities.

When the regular THz PC cavity is designed, two type of air holes are added in the center of the cavity and two novel THz PC cavities are formed as shown in Fig. 1(a). In type-1 THz PC cavity, ellipse air hole with longer radius $a$ and shorter radius $b$ is inserted into the center and the surface of the air hole is convex. While in type-2 THz PC cavity, two ellipse holes with longer radius $r_a$ and shorter radius $r_b$ are shifted in the opposite direction of the y-axis and an air hole with concave surface is formed. In view of y-direction, there is a low refractive index air slot sandwiched by high refractive index materials. This is the typical slot structure [17] which confines and guides electromagnetic wave in a nanometer-size low refractive index material due to the electric field discontinuity at the interface between high-index-contrast materials. The so-called slot effect can great enhance electromagnetic wave density in slot structures, especially for slot with concave surface [21, 29-31], bringing benefits to nonlinear processes [32], sensing [33], and nanolasing [30].

From Figs. 1(b) and 1(c), one can see that extremely electric field enhancements occur at the center of two novel proposed THz PC cavities. Throughout the paper, $r_b$ is set as 100 nm, i.e., the slot width is as large as 200 nm; $r_a$ is ranged from 100 nm to 4000 nm. From the experimental point of view, it is possible to prepare such large slot because sub-10 nm slot can be fabricated by atomic layer lithography technique [34]. The dependences of the most relevant characteristics, including quality factor ($Q$), mode volume ($V$) and resonant wavelength ($\lambda$), as functions of $r_a$ in both types of THz PC cavities are calculated and presented in Fig. 2. Here, we define that $r_a$ is positive in type-1 structure and negative in type-2 structure, and the two structures converge at $r_a= 0$. The calculated $Q$-factors are shown in Fig. 2(a), with the total $Q_{tot}$ determined by the material absorption-related $Q_{abs}$ and radiation-related $Q_{rad}$ as $1/Q_{tot} = 1/Q_{abs}$

+ $1/Q_{rad}$, and $Q_{rad}$ is calculated by ignoring the imagine part of the permittivity of Si. The stars in Fig. 2 indicates the regular cavities without nanoholes. Similar tendency is observed in Fig. 2 as $|r_a|$ increasing in the two THz PC cavities. In Fig. 2(a), $Q_{rad}$ decreases as $|r_a|$ increases in both two THz PC cavities, which might be attributed to the air slot induced coupling between the cavity resonance mode and free-space radiation modes. Since the field distribution of the cavity mode is rarely changed except the region close to the nanohole, the $Q_{abs}$ remains constant and $Q_{tot}$ decreases as $|r_a|$ increases in both two types of THz PC cavities. $Q_{tot}$ is firstly dominated by the absorption loss as $|r_a|$ increases and then dominated by the radiation loss when $|r_a|$ is larger than about 3000 nm.

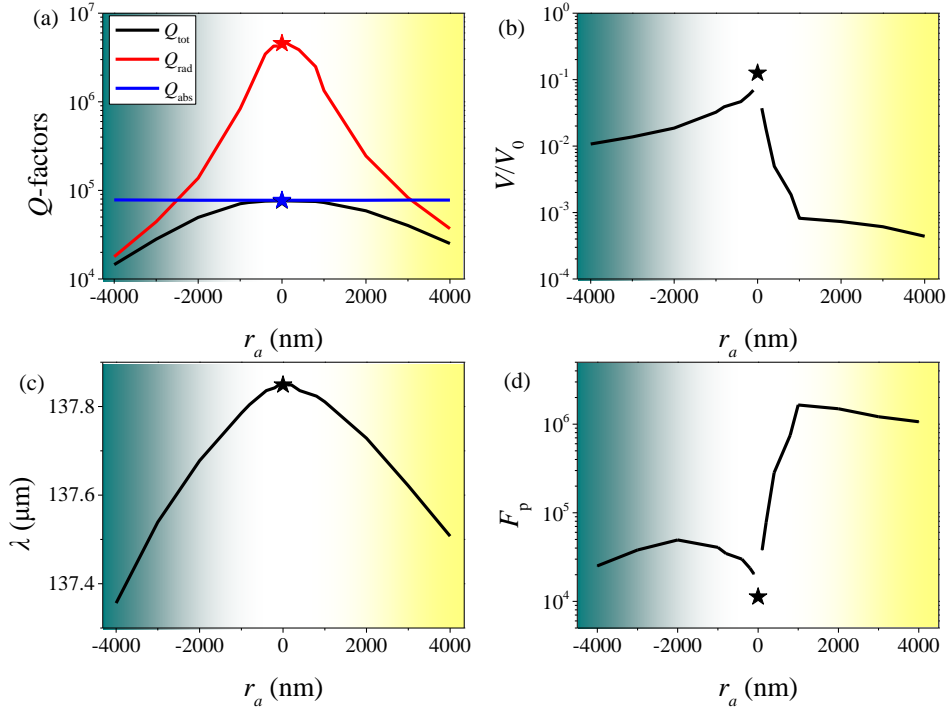

Fig. 2 Quality factors (a), normalized mode volume (b), resonant wavelength (c) and Purcell factor (d) versus $r_a$.

The mode volume $V$ is calculated by the ratio of the total electrical energy to the maximum electric energy density [20]:

$$V = \frac{\int_{all} \varepsilon(r)|E(r)|^2 d^3r}{\max\left[\varepsilon(r)|E(r)|^2\right]}. \tag{1}$$

The mode volume is normalized by $V_0$ (= $(\lambda/2)^3$), which represents the diffraction-limited mode volume in free space; $\lambda$ is the resonant wavelength (Fig. 2(c)). In Fig. 2(b), the normalized mode volume decreases dramatically as $|r_a|$ increases in both two THz PC cavities, which indicates that the field enhancement becomes stronger for larger $|r_a|$. In regular THz PC, the mode volume is about one order of magnitude smaller than $V_0$. In type-2 structure, the mode volume can be reduced to about two orders of magnitude smaller than the $V_0$. While in type-1 structure, the mode volume can be further squeezed to be more than three orders of magnitude smaller than $V_0$. The enhancement factor for mode volume $E_v$, which is defined as the ratio of the mode volume in the proposed THz PC cavities with that in the regular THz PC cavity, is plotted in Fig. S1 where indicates the $E_v$ in type-1 structure can reach to 288.

There exists a trade-off between the $Q$ factor and $V$, so another key parameter Purcell factor $F_p$ which represents the spontaneous emission rate enhancements factor of emitters is shown in Fig. 2(d). The Purcell factors is calculated by [20]:

$$F_p = \frac{3}{4\pi^2}\left(\frac{\lambda}{n}\right)^3\left(\frac{Q}{V}\right), \tag{2}$$

where $n$ is the material refractive index. The Purcell factors increase first then decrease as $|r_a|$ increases; the optimal values of $r_a$ are about 1000 nm and 2000 nm for type-1 and type-2 structures, respectively. In type-1 structure, strong enhancement of Purcell factor higher than $10^6$ is obtained, which is several orders larger than those in metal cavities [7, 8, 12] at THz Frequencies. As known, radiation emission rate is especially small in terahertz frequency range [7]. Thus, the proposed structures, especially for type-1 structure, provide an excellent platform for enhancing the spontaneous emission. In the following section, the underlying physical mechanisms of field enhancement in our proposed cavity structure is revealed.

### 3. Mechanism of extremely field enhancement: slot and anti-slot effect

In this section, we study the physical mechanism by the detailed field distribution around the nanohole. As shown by the field distribution in inset of Fig. 3(b), the main electric field is in the $y$-axis, the air hole serves as a traditional slot structure in view of $y$ direction. Electromagnetic field is highly confined in the low refractive index region due to so-called "slot effect" [17]. Theoretically, this field enhancement is arising from the electromagnetic boundary condition that the normal component of electric displacement field **D** should be continuous at the interface. As a result, the electric field **E** must undergo a large discontinuity with much higher amplitude in the low-index side. So, for point A, the normal components of **D** and **E** must satisfy

$$D_{A,h} = D_{A,l}, \tag{3}$$

$$E_{A,l} = \frac{\varepsilon_h}{\varepsilon_l}E_{A,h}, \tag{4}$$

where $\varepsilon_h$ and $\varepsilon_l$ are the permittivities, with the subscripts $h$, $l$ denotes the high and low refractive index materials, respectively. Equations (3) and (4) are confirmed by the numerical results of the normal components of **D** and **E** (the results are shown in Figs. S2). The electric displacement field **D** is continuous at the interface of silicon and air, such as point A in the inset of Fig. 3(b), while the electric field **E** is greatly enhanced in the air slot. The amplitudes of normal components of electric field **E**, i.e., |E|, at the interface of silicon and air are $E_{A,h} = 7.26 \times 10^5$ a.u. (arbitrary unit) and $E_{A,l} = 7.72 \times 10^6$ a.u.. The enhancement factor is $E_{A,l}/E_{A,h} \sim 10.6$, which agrees well with the theoretically value of $\varepsilon_h/\varepsilon_l \sim 11.7$ as predicted by Eq. (4). As a result, the electric filed energy density $W = \bm{DE}/2$ in the air slot is also enhanced (inset of inset of Fig. 3(c)).

At point B in the inset of Fig. 3(b), the tangential of the electric field **E** is continuous according to the electromagnetic boundary condition; the tangential components of **D** and **E** must satisfy

$$E_{B,h} = E_{B,l}, \tag{5}$$

$$D_{B,h} = \frac{\varepsilon_h}{\varepsilon_l}D_{B,l}. \tag{6}$$

Thus, the tangential components of **D** and the electric field energy density $W$ are enhanced in the high refractive index materials which is different from the traditional slot effect. This effect is called "anti-slot effect". Calculation results verify that $|E|(x, 0)$ is continuous and $|D|(x, 0)$ is discontinuous at the interface of the silicon and air. $|D|_{B,h}$ and $|D|_{B,l}$ at point B are $8\times 10^{-4}$ a.u. and $6.75\times 10^{-5}$ a.u., respectively. The enhancement factor of $|D|_{B,h}/|D|_{B,l}$ is about 11.9 which also agrees well with the theoretically value of $\varepsilon_h/\varepsilon_l \sim 11.7$.

The mechanism of the field enhancement can be summarized as: the electric filed energy density $W$ is enhanced in the air slot by a factor of $\varepsilon_h/\varepsilon_l$ due to the slot effect, then it is enhanced further in the high index region by a factor of $\varepsilon_h/\varepsilon_l$ due to the anti-slot effect. The electric filed energy density $W$ is finally enhanced by a factor of $(\varepsilon_h/\varepsilon_l)^2$ as shown in Fig. 3. Importantly, $W$ is maximized in the high refractive index materials, which is different from the traditional slot structures [17, 35]. This is very helpful, especially at THz frequencies where the radiation emission rate is limited, for enhancing light and matter interactions by embedding emitters in high refractive index materials.

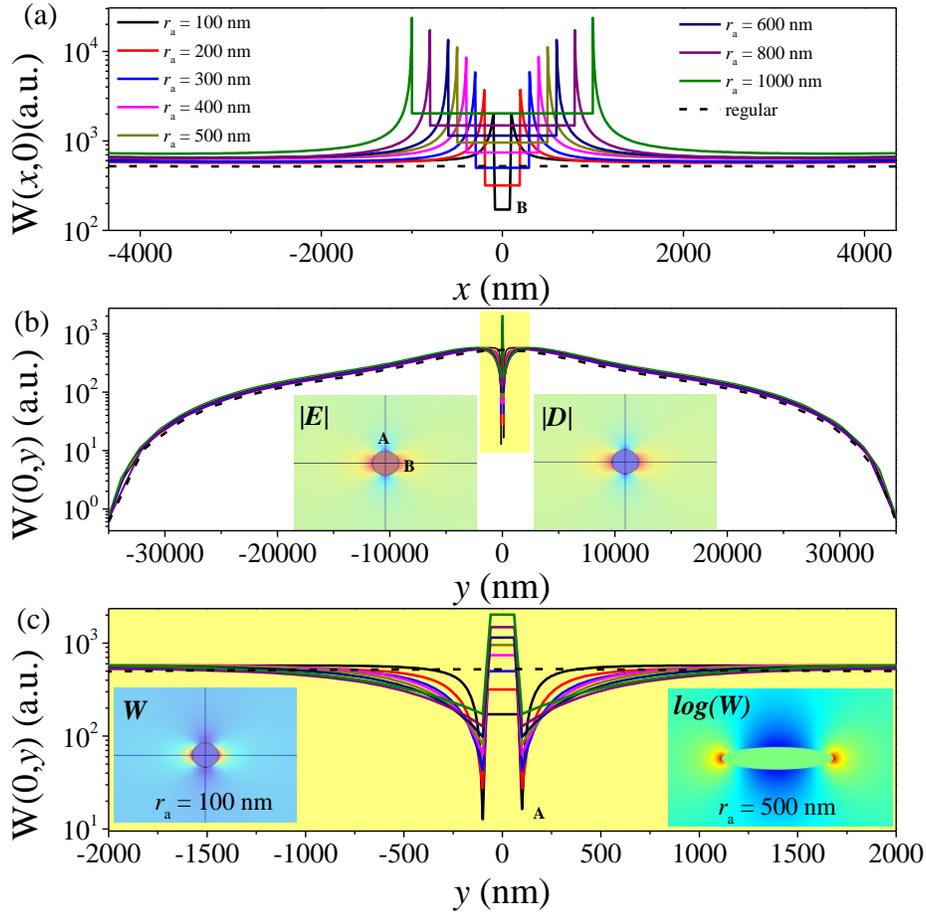

Fig. 3 (a) $W(x, 0)$-field, (b) $W(0, y)$-field and (c) enlarged view of $W(0, y)$-field distributions in type-1 structure for different $r_a$. Insets in (b) are the 2-D $|E|$ and $|D|$ distributions. Insets in (c) are the 2-D $W$ distributions with $r_a = 100$ nm and $r_a = 500$ nm, respectively.

As shown in Fig. 3, the electric energy density $W$ is always uniform in the air hole region in type-1 structure. For $r_a < 300$ nm, the $W$ in the air hole is smaller than that in regular PC cavity (dashed lines in Fig. 3). The $W$ is always maximized in the $x$ direction attributed to the anti-slot effect. Here, without loss of generality and saving calculation time, two-dimensional model is used for calculating the filed distribution, as well as the calculations in Figs. S3-S5.

For larger $r_a$, local field enhancement around the two corners of ellipse air hole can be seen intuitively from the right inset in Fig. 3(c). The $W$ increases and becomes more localized as $r_a$ increases, but the enhancement of $W$ in point A is always kept to about $(\varepsilon_h/\varepsilon_l)^2$. This local field enhancement in type-1 structure looks a bit similar with that in tip or bowtie structures [21-23]. However, the essential different is that, in type-1 structure, the electric energy field is enhanced in the high refractive index materials. While in conventional tip or bowtie structures, the electric energy field is enhanced in the low refractive index materials.

Similar in type-1 structure, in type-2 THz PC cavities, |***E***|-field and $W(x, 0)$-field are enhanced by the slot effect in $y$ direction, |***D***|-field and $W(0, y)$-field are enhanced by the anti-slot effect in $x$ direction. Modal field distributions in type-2 THz PC cavities with $r_a$= 100 nm can be found in Fig. S3. However, the mode volumes in type-1 structure is lower than that in type-1 structure at the same $|r_a|$, especially for larger $|r_a|$.

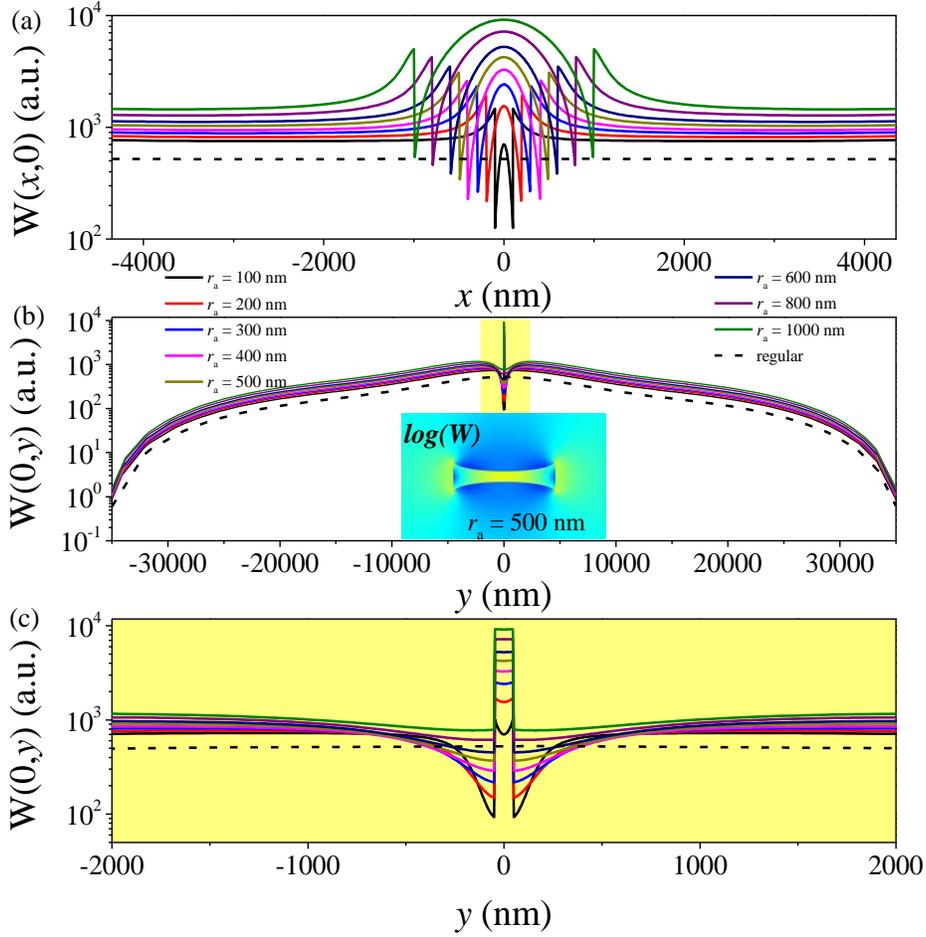

Fig. 4 (a)$W(x, 0)$-field, (b)$W(0, y)$-field and (c) enlarged view of $W(0, y)$-field distributions in type-1 structure. Inset in (b) is the 2-D $W$ distributions with $r_a$ = 500 nm.

To reveal this, we plot the electric filed energy density distributions in type-2 structure with different $|r_a|$ as comparison in Fig. 4, where shows the distribution of $W$ when $r_a$ varies from 100 nm to 1000 nm. In type-2 structure, the electric energy density $W$ is not uniform in the air hole region due to that the surface of air hole is concave. For $|r_a| < 300$ nm, the $W$ is maximized in the interface of the silicon and air along the $x$ direction (that is Point B) which can be seen form Fig. 4(a). While for $|r_a| > 300$ nm, the $W$ is maximized in the interface of the silicon and air along the y direction (that is Point A) which can be seen form Figs. 4(b) and 4(c). That indicates the anti-slot effect is dominant in the enhancement of $W$ when $|r_a| < 300$ nm, the slot effect is dominant in the enhancement of $W$ when $|r_a| > 300$ nm. However, the slot effect enhances the $W$ in the whole slot region as shown in the inset of Fig. 4(b). So the $W$ is relatively spread as compared with that in type-1 structure where the $W$ is maximized and localized the two corners (point B) along the $x$ direction which can be seen in Fig. 3. Besides, The $W$ is not finally enhanced by a factor of $(\varepsilon_h/\varepsilon_l)^2$ at Point B (interface of silicon and air for $y = 0$) in type-2 structure, because the $|E|$-field is not enhanced by a factor of $\varepsilon_h/\varepsilon_l$ uniformly in the whole air hole (Fig. S3). As a result, the mode volume in type-1 structure is smaller than that in type-2 structure at the same $r_a$. It is worth noting that there are four points where the fields cannot be converged in calculation due to its sharp corners as shown in the inset in Fig. 4(b). So, the maximum of the electric energy field in these points is excluded and we select the maximum in $x$ direction ($y = 0$) or in $y$ direction ($x = 0$) of the electric energy field when calculating the mode volumes.

## 4. Horizontally coupled air hole array

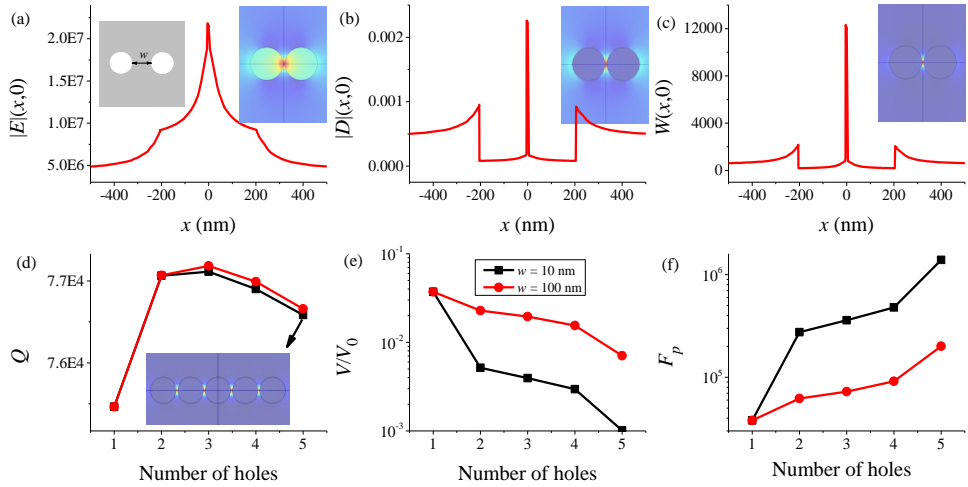

Fig. 5 (a)-(c) show the 1-D $|E|$-field, $|D|$-field and $W$-field distributions in the $x$ direction in two horizontally air holes coupled THz PC cavities. (d)-(f) show the dependence of $Q$-factor, $V/V_0$ and $F_p$ as a function of the number of holes.

Finally, we also propose another strategy to further enhance the THz waves by cascading type-1 structures, where a chain of horizontally coupled circular air hole array is inserted into the center of the THz PC cavities. Figures 5(a)-5(c) show the 1-D $|E|$-field, $|D|$-field and $W$-field distributions in the $x$ direction in THz PC cavities with two horizontally air holes separated by distance $w$ (insets in Fig. 5(a)). $|E|$-field is continuous at the outer interface, i.e., $x = \pm 205$ nm, of the silicon and air as shown in Fig. 5(a) for $w = 10$ nm. While huge enhancement of electric field occurs at the silicon ridge between the two air holes. So, huge enhancement of electric displacement field and energy field also occur at the silicon ridge as shown in Figs. 5(b) and 5(c) (see Fig. S5 for more air holes condition). The THz wave is highly confined in the nanoscale silicon ridge which can be seen in the inset of Fig. 5(c). The mode volume decreases by one order magnitude as compared with type-1 structure with one circular air hole (Fig. 5(e)). Accompanied with the squeeze of electromagnetic field, radiation loss is also suppressed and the $Q$ factor increases (Fig. 5(d)). According to the simulation results for more horizontally aligned air holes and larger silicon ridges, we found that the enhancement increases with the number of holes, while the enhancement reduces with larger silicon ridge width. When five air holes coupled with $w = 10$ nm, the mode volume is three orders smaller than the diffraction limited mode volume $V_0$, which leads to a huge $F_p$ larger than $10^6$.

## 5. Summary

We have proposed a novel type of all-dielectric THz PC cavity, which can extremely concentrate the electric field of THz waves by inserting air holes into the center of the cavity. By utilizing the boundary conditions for electromagnetic waves at dielectric interfaces, the electric field can be firstly enhanced by a factor of $\varepsilon_h/\varepsilon_l$ via the slot effect, and the electric displacement field can be further enhanced by a factor of $\varepsilon_h/\varepsilon_l$ via anti-slot effect. Then, the electric energy density is enhanced by a factor of $(\varepsilon_h/\varepsilon_l)^2$, and the mode volume of the cavity can be reduced to be three orders of magnitude smaller than the diffraction-limited mode volume. The mode volume can be enhanced by a factor of 288 as compared that in regular THz PC cavity, while a high $Q$-factor above $10^4$ can also be maintained. Such highly localized field by our designed cavity structure can greatly enhance the light-matter interactions in the THz frequency band, which is highly desired for applications ranging from sensing [36, 37] to the coherent THz sources [5, 6, 38].

**Funding**

National Key Basic Research Program of China (973 project) under Grant No. 2015CB352006; National Natural Science Foundation of China under Grants Nos. 61705039, 61505195, 91536219 and 61335011; China Postdoctoral Science Foundation (2017M610389); Fujian Provincial Program for Distinguished Young Scientists in University; Fujian Provincial Key Project of Natural Science Foundation for Young Scientists in University (JZ160423); Program for Changjiang Scholars and Innovative Research Team in University under Grant No. IRT_15R10; Special Funds of the Central Government Guiding Local Science and Technology Development (2017L3009); CLZ is supported by Anhui Initiative in Quantum Information Technologies (AHY130000).